\documentclass[acmlarge]{acmart}
\pdfoutput=1
\AtBeginDocument{%
  }

\setcopyright{none}
\renewcommand\footnotetextcopyrightpermission[1]{} 

\acmConference[ICMLT 2023]{2023 8th International Conference on Machine Learning Technologies (ICMLT)}{March 10--12 2023}{Stockholm, Sweden}
\acmPrice{15.00}
\acmISBN{978-1-4503-XXXX-X/18/06}

\usepackage{graphicx}

\newcommand\HEMAN{\mbox{HE-MAN}}
\newcommand\HEMANconcrete{\HEMAN{}-Concrete}
\newcommand\HEMANtenseal{\HEMAN{}-TenSEAL}
\newcommand\toolset{toolset}

\newcommand\concrete{\mbox{Concrete}}
\newcommand\meanstd{mean-std}

\usepackage{url}
\usepackage{tikz}
\usepackage{tikzsymbols}
\usetikzlibrary{positioning}
\usetikzlibrary{calc,through}
\usepackage{multirow}
\usepackage{booktabs} 
\usepackage{amsmath}
\usepackage{arydshln} 
\usepackage{color, colortbl} 
\usepackage[htt]{hyphenat}
\usepackage{fontawesome}
\usepackage{listings}
\usepackage{enumitem}
\usepackage{caption}
\usepackage{wrapfig}

\usepackage{pgfplots}
\pgfplotsset{
    compat=1.17,
    legend image code/.code={
        \draw plot coordinates {
            (0cm,0cm)
            (0.75cm,0cm)        
            (0.75cm,0cm) 
        };
    }
}

\definecolor{gray}{gray}{0.9}
\newcommand{\circlefilled}{\tikz\draw[fill] (0,0) circle (0.1);}
\newcommand{\circleempty}{\tikz\draw (0,0) circle (0.1);}

\begin{document}
\pagestyle{plain}

\title{HE-MAN -- Homomorphically Encrypted MAchine learning with oNnx models}

\author{Martin Nocker}
\authornote{Correspondence: martin.nocker@mci.edu, MCI~Management~Center~Innsbruck, Universitätsstraße 15, 6020 Innsbruck, Austria}
\affiliation{%
  \institution{MCI~Management~Center~Innsbruck, Innsbruck}
  \city{Innsbruck}
  \country{Austria}
  \postcode{6020}
}
\email{martin.nocker@mci.edu}

\author{David~Drexel}
\affiliation{%
  \institution{MCI Management Center Innsbruck, Innsbruck}
  \city{Innsbruck}
  \country{Austria}
  \postcode{6020}
}
\email{david.drexel@mci.edu}

\author{Michael Rader}
\affiliation{%
  \institution{Fraunhofer~Austria~Research~GmbH, Wattens}
  \city{Wattens}
  \country{Austria}}
\email{michael.rader@fraunhofer.at}

\author{Alessio Montuoro}
\affiliation{%
  \institution{SCCH~Software~Competence Center Hagenberg, Hagenberg}
  \city{Hagenberg}
  \country{Austria}
}
\email{alessio.montuoro@scch.at}

\author{Pascal~Schöttle}
\affiliation{%
  \institution{MCI~Management~Center~Innsbruck, Innsbruck}
  \city{Innsbruck}
  \country{Austria}
  \postcode{6020}
}
\email{pascal.schoettle@mci.edu}

\renewcommand{\shortauthors}{Martin Nocker et al.}

\begin{abstract}
Machine learning (ML) algorithms are increasingly important for the success of products and services, especially considering the growing amount and availability of data. This also holds for areas handling sensitive data, e.g. applications processing medical data or facial images. However, people are reluctant to pass their personal sensitive data to a ML service provider. At the same time, service providers have a strong interest in protecting their intellectual property and therefore refrain from publicly sharing their ML model. Fully homomorphic encryption (FHE) is a promising technique to enable individuals using ML services without giving up privacy and protecting the ML model of service providers at the same time. 
Despite steady improvements, FHE is still hardly integrated in today's ML applications. Reasons for that are, among others, that existing implementations either require the user to possess expertise in FHE, do not feature an easy ML framework integration, or have to approximate non-polynomial activations.

We introduce \HEMAN{}, an open-source two-party machine learning \toolset{} for privacy preserving inference with ONNX models and homomorphically encrypted data. Both the model and the input data do not have to be disclosed. \HEMAN{} abstracts cryptographic details away from the users, thus expertise in FHE is not required for either party. \HEMAN 's security relies on its underlying FHE schemes. For now, we integrate two different homomorphic encryption schemes, namely \concrete{} and TenSEAL.
Compared to prior work, \HEMAN{} supports a broad range of ML models in ONNX format out of the box without sacrificing accuracy.

We evaluate the performance of our implementation on different network architectures classifying handwritten digits and performing face recognition and report accuracy and latency of the homomorphically encrypted inference. Cryptographic parameters are automatically derived by the tools. We show that the accuracy of \HEMAN{} is on par with models using plaintext input while inference latency is several orders of magnitude higher compared to the plaintext case.
\end{abstract}

\ccsdesc[300]{Security and privacy~Cryptography}
\ccsdesc[300]{Security and privacy~Privacy protections}
\ccsdesc[300]{Security and privacy~Public key encryption}
\ccsdesc[300]{Computing methodologies~Neural networks}

\keywords{Homomorphic Encryption, Machine Learning as a Service, Secure and Privacy-Preserving Machine Learning}

\maketitle

\section{Introduction}
Today's products and services increasingly benefit from the integration of evermore powerful machine learning (ML) algorithms. Furthermore, the amount and availability of data is steadily increasing and penetrates all kinds of areas of our daily lives. Thus, it comes as no surprise that also more and more personal data is collected and society as a whole could potentially benefit from the usage of this data in machine learning contexts. But, by definition, personal data is sensitive, thus, people are understandably reluctant to send this kind of data to, e.g. ML service providers. On the other hand, providers of Machine Learning as a Service (MLaaS) do not want to share their models as these constitute their intellectual property. One solution to this dead end where neither party benefits from the advantages of modern ML approaches is the combination of ML with fully homomorphic encryption (FHE). In comparison to ``standard'' encryption schemes, homomorphic encryption (HE) schemes allow for computations on encrypted data without the need for decryption in between.

Despite the ongoing development and several breakthroughs, making FHE ever more efficient, FHE is still hardly integrated into ML applications. Besides a non-negligible performance-overhead compared to computations on cleartext data, this also originates from FHE programs being hard to implement as expertise in software development with FHE and FHE in general is necessary. Therefore, there is a strong need in the field to ease the integration of FHE into ML applications for non-cryptographers.

In this paper, we present \HEMAN{}, a \toolset{} consisting of HE-MAN-Concrete and \HEMANtenseal{}, two tools based on the open-source FHE libraries \concrete{}~\cite{chillotti2020concrete} and TenSEAL~\cite{tenseal2021}, respectively.
The aim of \HEMAN{} is to enable homomorphic inference using cleartext models, provided in the Open Neural Network Exchange (ONNX) format~\cite{bai2019onnx}, for homomorphically encrypted inputs, while preserving the privacy of the model and the input data.
Figure~\ref{fig:firstlady} illustrates the high-level structure of \HEMAN{}. The model owner offers a neural network to be used for inference on encrypted inputs from external clients in return for a reward per executed inference. The model is never passed to the data owner, while input data is processed in encrypted form only. Neither party gains knowledge about the other party's sensitive information.
The reward transaction is not part of the current \HEMAN{} implementation.

As implementing FHE programs requires expertise in FHE and is a tedious task, \HEMAN{} abstracts cryptographic details away from the user. Encryption parameters for secure and accurate homomorphic computations are automatically derived. Operations are performed efficiently and are replaced by FHE-friendly operations when necessary.
\\\\
\textbf{Contributions:}
\HEMAN{} is, to the best of our knowledge, the first work that combines neural network inference with \concrete{} and TenSEAL under a common framework.  
\concrete{} features programmable bootstrapping, enabling arbitrary activation functions, thus, being highly suitable for neural network computation~\cite{chillotti2020programmable}.
The support of TenSEAL, a Python library built on top of Microsoft SEAL~\cite{sealcrypto}, has the advantage 
of encrypting and computing on multiple values in one ciphertext, known as batching.

To summarize, we make the following contributions:
\begin{itemize}[noitemsep,topsep=0pt]
    \item We present \HEMAN, an open-source two-party machine learning \toolset{}, that preserves the privacy of both, ML model and inference input.  
    \item Through the usage of ONNX models, \HEMAN{} supports a broad range of pretrained models independent of ML frameworks.
    \item \HEMAN{} exhibits a user-friendly command line interface, where neither model owner nor data owner needs expertise in fully homomorphic encryption.
    \item For now, \HEMAN{} supports the two popular FHE libraries \concrete{} and TenSEAL, but is easily extendable to other libraries.
    \item The usage of \HEMANconcrete{} enables the computation of nonlinear activations, e.g. ReLU, without polynomial approximation.
    \item Our implementation builds on top of TenSEAL such that it supports an arbitrary number of convolutions at an arbitrary layer in the respective network.
\end{itemize}
\ \\
\textbf{Organization:} The rest of this paper is organized as follows. Section~\ref{sec:background} introduces homomorphic encryption and FHE schemes integrated in \HEMAN{}. Next, section~\ref{sec:threat-model} addresses the threat model. Sections~\ref{sec:he-man} and~\ref{sec:designchoices} present the protocol and design choices of \HEMAN, respectively before experimental evaluations are discussed in section~\ref{sec:evaluation}. Related work is covered in section~\ref{sec:related-work} before section~\ref{sec:conclusion} concludes this paper.

\begin{table}
\centering
\parbox{.42\linewidth}{
    \centering
    \begin{tikzpicture}[
    scale=0.925,
    >=latex,
    every node/.style={transform shape, font=\sffamily},
    neuron/.style={circle, inner sep=0pt, draw, minimum size=1.5mm, thick },
    arrowstyle/.style={line width=1pt},
  ]
  
  \node[rectangle,draw,very thick,minimum height=20mm,minimum width=20mm] at (0,0) (heman) {HE-MAN};
  
  \node at ($(heman)+(3,0.5)$) (input) {\faFileO};
  \node at ($(heman)+(3,-0.5)$) (output) {\faFileTextO};
  \draw[->,arrowstyle] (input) -- node[above=-3pt] {\footnotesize$\begin{array}{c}\mbox{encrypted}\\\mbox{input}\end{array}$} (input-|heman.east);
  \draw[->,arrowstyle] (output-|heman.east) -- node[above=-3pt] {\footnotesize$\begin{array}{c}\mbox{encrypted}\\\mbox{result}\end{array}$} (output);
  \node[above=5mm of input, align=center] (dolabel) {\small Data\\\small Owner};
  
  \node[minimum width=10mm] at ($(heman)+(-3.25,0.5)$) (model) {};
  \node at ($(heman)+(-3.25,-0.5)$) (reward) {\faDollar};
  \draw[->,arrowstyle] (model) -- node[above] {\footnotesize model} (model-|heman.west);
  \draw[->,arrowstyle] (reward-|heman.west) -- node[above] {\footnotesize reward} (reward-|model.east);
  \node[align=center] at (model|-dolabel) {\small Model\\\small Owner};
  
  \node[neuron] (n11) at (model) {};
  \node[neuron] (n10) at ($(n11)+(0,0.3)$) {};
  \node[neuron] (n12) at ($(n11)+(0,-0.3)$) {};
  \node[neuron] (n00) at ($(n11)+(-0.3,0.15)$) {};
  \node[neuron] (n01) at ($(n00)+(0,-0.3)$) {};
  \node[neuron] (n20) at ($(n11)+(0.3,0.15)$) {};
  \node[neuron] (n21) at ($(n20)+(0,-0.3)$) {};
  
  \foreach \a in {0,1} {
    \draw[-] (n0\a) -- ++(-0.15,0);
    \foreach \b in {0,1,2}
      \draw[-] (n0\a) -- (n1\b);
  }
  \foreach \a in {0,1} {
    \draw[-] (n2\a) -- ++(0.15,0);
    \foreach \b in {0,1,2}
      \draw[-] (n1\b) -- (n2\a);
  }
\end{tikzpicture}
    \captionof{figure}{Involved parties and their respective inputs and outputs when using \HEMAN{} for privacy-preserving machine learning.}
    \label{fig:firstlady}
}
\qquad\qquad
\parbox{.42\linewidth}{
\caption{State-of-the-art FHE libraries and the schemes they implement}
\resizebox{\linewidth}{!}{%
\centering
\begin{tabular}{lccc}
        \toprule
        \multirow{2}{*}{Library} & \multicolumn{3}{c}{Schemes}\\ \cmidrule(lr){2-4}
         & BFV~\cite{fan2012somewhat} & CKKS~\cite{cheon2017homomorphic} & TFHE/CGGI~\cite{chillotti2016faster}
        \\ \midrule
        \concrete{}~\cite{chillotti2020concrete} & \circleempty & \circleempty & \circlefilled \\
        HElib~\cite{halevi2018faster} & \circlefilled & \circlefilled & \circleempty \\
        lattigo~\cite{mouchet2020lattigo} & \circlefilled & \circlefilled & \circleempty \\
        OpenFHE~\cite{OpenFHE} & \circlefilled & \circlefilled & \circlefilled \\
        SEAL~\cite{sealcrypto} & \circlefilled & \circlefilled & \circleempty \\
        \bottomrule
    \end{tabular}\label{tab:libraries}}
}
\end{table}

\section{Background}\label{sec:background}
\subsection{Homomorphic Encryption}
Homomorphic encryption schemes allow for computations on encrypted data without decryption, i.e. besides the encryption and decryption functions $\mathbb{E}$ and $\mathbb{D}$, there exist operators $\oplus$ and $\otimes$ such that for any plaintext elements $x_1, x_2$ in the plaintext space
$\mathbb{D}(\mathbb{E}(x_1)\oplus\mathbb{E}(x_2)) = x_1 + x_2$ and $\mathbb{D}(\mathbb{E}(x_1)\otimes\mathbb{E}(x_2)) = x_1 \times x_2$,
where $+$ and $\times$ represent addition and multiplication in the plaintext space.
A fully homomorphic encryption (FHE) scheme supports both operations.
Long before the first FHE scheme, so-called partially homomorphic encryption schemes (PHE) have been developed. Here, only one operation, e.g. addition in the Paillier scheme~\cite{paillier1999public} or multiplication in textbook RSA~\cite{rivest1978method}, is supported.
First proposed in~\cite{rivest1978data} as \emph{privacy homomorphism}, it was Gentry's breakthrough to construct the first fully homomorphic encryption scheme~\cite{gentry2009fully} that started the development of subsequent improved FHE schemes.

The ring learning with errors (RLWE) hardness assumption~\cite{lyubashevsky2010ideal} is the basis of modern FHE schemes' security, i.e. during encryption random noise is added to the ciphertext. Performing operations on RLWE-ciphertexts increases the noise level. Additions increase the noise level negligible, but multiplications add a significant amount of noise to the ciphertext. Decryption will only return a correct result, if the noise level stays below a certain noise threshold. Therefore, the multiplicative depth of a homomorphic computation is often the limiting factor in theses schemes.
An additional operation called \emph{bootstrapping}, which was introduced by Gentry~\cite{gentry2009fully}, reduces the noise level of a ciphertext by homomorphically evaluating the decryption circuit of its FHE scheme and re-encrypting the ciphertext. However, this operation is computationally very expensive and therefore avoided as much as possible. \emph{Leveled} schemes can be used to avoid bootstrapping completely~\cite{brakerski2014leveled,fan2012somewhat,cheon2017homomorphic}. Here, the encryption parameters are set large enough, such that the entire computation can be completed without bootstrapping.
Yet, it is not possible to increase the parameters arbitrarily, as increasing parameters comes at the cost of decreased computational performance.

\subsection{FHE schemes}
Table~\ref{tab:libraries} gives an overview of state-of-the-art FHE libraries and their supported schemes. Implemented schemes are denoted by \circlefilled, missing schemes by \circleempty.
The security of FHE schemes is specified by the security level $\lambda$, with \mbox{$\lambda=128$}~bits implying approximately $2^{128}$ operations are required to break the encryption. $\lambda$ is a function of the encryption parameters which heavily affect the runtime performance.

\HEMAN{} utilizes the TFHE\footnote{also called CGGI after the authors' initials}~\cite{chillotti2016faster} scheme implemented in version~0.1 of the \concrete{}~\cite{chillotti2020concrete} library and the leveled implementation of the CKKS~\cite{cheon2017homomorphic} scheme in the TenSEAL~\cite{tenseal2021} library version~0.3.12 which is a Python wrapper around the Microsoft Simple Arithmetic Library (SEAL)~\cite{sealcrypto} version~4.0.
\\ \\
\textbf{TFHE: }
Subsequent improvements extend TFHE to arithmetic circuits and introduce \emph{Programmable Bootstrapping (PBS)}~\cite{chillotti2020programmable} which works as a look-up table during bootstrapping. This enables the homomorphic evaluation of arbitrary univariate functions to a ciphertext.
The cryptographic parameters of interest for TFHE are the dimension of vectors of polynomials $N$, the vector length $k$ and the standard deviation $\sigma$ of gaussian noise used for encryption. It can roughly be said that higher values of $\sigma$ increase the security of the calculation, but degrade the accuracy of the result after decryption as the noise becomes more significant. Higher choices of $k$ and $N$ in turn allow lowering $\sigma$ while keeping the same amount of security, at the cost of increased computational effort.
\\ \\
\textbf{CKKS: }
CKKS ciphertexts are degree-$N$ polynomials with integer coefficients mod $q$, where $N$ is the \emph{polynomial modulus degree} and a power of two and $q$ is the \emph{coefficient modulus}. A larger polynomial modulus degree yields larger ciphertexts and slower operation execution, but enables homomorphic computations with higher multiplicative depth. CKKS supports computations on floating-point numbers by using fixed-point arithmetic.


CKKS features a technique called \emph{batching} which allows to pack a vector of $N/2$ complex numbers into a single ciphertext. The number of encrypted values $N/2$ within one ciphertext is also referred to as \emph{slots}. Operations performed on a ciphertext are evaluated element-wise on all encrypted values simultaneously, which allows for more efficient implementations. However, increasing $N$ decreases computational performance~\cite{laine2017simple}.

\begin{figure*} 
    \centering
    \begin{tikzpicture}[
  scale=0.925,
  >=latex,
  every node/.style={transform shape, font={\footnotesize\sffamily}},
  neuron/.style={circle, inner sep=0pt, draw, minimum size=1.5mm, thick },
  arrstyle/.style={line width=1.25pt, color=black!75},
  toolnode/.style={draw,rectangle,inner sep=1mm,thick,minimum height=5mm,minimum width=22mm, font=\small},
  toolnodeoff/.style={toolnode,color=black!25},
  file/.style={draw,fill=white,text=black,line width=0.5pt,thick},
  clientfile/.style={file,minimum width=22mm,minimum height=5mm},
  publicfile/.style={file,minimum width=21mm},
  midwayfile/.style={midway,publicfile},
  stepnum/.style={draw,circle,inner sep=0.5mm,fill=white,thick}
  ]
  
  \node[rectangle,draw,very thick,minimum height=50mm,minimum width=30mm] at (4,2.5) (itool) {};
  \node[rectangle,draw,thick, minimum width=30mm,anchor=north] (itool-label) at (itool.north) {\normalsize HE-MAN};
  \node[toolnode,below=2mm of itool-label] (keyspec) {\texttt{KeyParams} \faKey};
  \node[toolnodeoff,below=10mm of itool-label] (keygen) {\texttt{KeyGen} \faKey};
  \node[toolnodeoff,below=18mm of itool-label] (encrypt) {\texttt{Encrypt} \faLock};
  \node[toolnode,below=26mm of itool-label,minimum height=8mm] (inf) {\texttt{Inference}};
  \node[toolnodeoff,below=37mm of itool-label] (decrypt) {\texttt{Decrypt} \faUnlockAlt};
  
  \node[rectangle,draw,very thick,minimum height=50mm,minimum width=30mm, fill=white] at ($(itool)+(7.5,0)$) (itool-c) {};
  \node[rectangle,draw,thick, minimum width=30mm,anchor=north] (itool-label-c) at (itool-c.north) {\normalsize HE-MAN};
  \node[toolnodeoff,below=2mm of itool-label-c] (keyspec-c) {\texttt{KeyParams} \faKey};
  \node[toolnode,below=10mm of itool-label-c] (keygen-c) {\texttt{KeyGen} \faKey};
  \node[toolnode,below=18mm of itool-label-c] (encrypt-c) {\texttt{Encrypt} \faLock};
  \node[toolnodeoff,below=26mm of itool-label-c,minimum height=8mm] (inf-c) {\texttt{Inference}};
  \node[toolnode,below=37mm of itool-label-c] (decrypt-c) {\texttt{Decrypt} \faUnlockAlt};
  
  \node[clientfile] at ($(keygen)-(3.25,0)$) (calibration) {$\begin{array}{c}\mbox{Calibration}\\\mbox{Data}\end{array}$\;\faFilesO\;};
  \draw[->,arrstyle] (calibration) -- (keyspec.181);
  
  \node (nn-center) at ($(itool-label)-(3.25,0.15)$) {};
  \node[neuron] at ($(nn-center)+(-0.3,0)$) (n00) {};
  \node[neuron] at ($(n00)+(0,0.2)$) (n01) {};
  \node[neuron] at ($(n00)-(0,0.2)$) (n02) {};
  \node[neuron] at ($(nn-center)+(0,0.1)$) (n10) {};
  \node[neuron] at ($(n10)+(0,0.2)$) (n11) {};
  \node[neuron] at ($(n10)-(0,0.2)$) (n12) {};
  \node[neuron] at ($(n10)-(0,0.4)$) (n13) {};
  \node[neuron] at ($(nn-center)+(0.3,0)$) (n20) {};
  \node[neuron] at ($(n20)+(0,0.2)$) (n21) {};
  \node[neuron] at ($(n20)-(0,0.2)$) (n22) {};
  \foreach \a in {0,1,2} {
    \draw[-,very thin] (n0\a) -- ++(-0.15,0);
    \foreach \b in {0,...,3}
      \draw[-,very thin] (n0\a) -- (n1\b);
  }
  \foreach \a in {0,...,2} {
    \draw[-,very thin] (n2\a) -- ++(0.15,0);
    \foreach \b in {0,...,3}
      \draw[-,very thin] (n1\b) -- (n2\a);
  }
  \node[anchor=south,outer sep=0.5mm] at (n11.north) (onnx-label) {ONNX-Model};
  \node[draw,thick,rectangle,minimum width=22mm, minimum height=15mm,anchor=north] at (onnx-label.north) (frame) {};
  \draw[->,arrstyle] (frame) -- (keyspec.170);
  
  \node[clientfile] at ($(inf)-(3.25,0)$) (cal-model) {$\begin{array}{c}\mbox{Calibrated}\\\mbox{ONNX-Model}\end{array}$};
  \draw[->,arrstyle] (cal-model) -- (inf);
  \draw[->,arrstyle] (keyspec.south west) -- (cal-model.north east);
  
   \node[clientfile] at ($(keygen-c)+(3.25,0)$) (keys) {\faKey$\begin{array}{c}\mbox{Secret Key}\end{array}$};
   \draw[->,arrstyle] (keygen-c) -- (keys);
   \draw[->,arrstyle] (keys.190) -- (encrypt-c.10);
   \draw[->,arrstyle] (keys.190) -- (decrypt-c.10);
   
   \node[clientfile] at (keys|-encrypt-c) (datain) {\faFileO$\begin{array}{c}\mbox{Input}\end{array}$};
   \draw[->,arrstyle] (datain) -- (encrypt-c);
   
   \node[clientfile] at (keys|-decrypt-c) (dataout) {\faFileTextO$\begin{array}{c}\mbox{Result}\end{array}$};
   \draw[->,arrstyle] (decrypt-c) -- (dataout);
  
  \draw[->,arrstyle] (keyspec) -- (keygen-c) node (encparams) [midwayfile] {$\begin{array}{c}\mbox{Parameters}\end{array}$\faKey};
  
  \draw[->,arrstyle] (keygen-c.185) -- (inf.17) node [above=-6pt,midwayfile] {$\begin{array}{c}\mbox{Evaluation}\\\mbox{Key}\end{array}$\faKey};
  
  \draw[->,arrstyle] (encrypt-c.west) -- (inf.east) node [below=-4.5pt,midwayfile] {$\begin{array}{c}\mbox{Encrypted}\\\mbox{Input}\end{array}$\;\faFile};
  
  \draw[->,arrstyle] (inf.350) -- (decrypt-c.west) node [below=-7pt, midwayfile] {$\begin{array}{c}\mbox{Encrypted}\\\mbox{Result}\end{array}$\;\faFile};
  
  \node[publicfile, above=2mm of encparams, inner sep=0.3mm, minimum height=8mm] (seclevel) {$\begin{array}{c}\mbox{Security}\\\mbox{Level}\end{array}$\large$\lambda$};
  
  \draw[->,arrstyle] (seclevel) -- (keyspec);

  \draw[loosely dashed] ($(itool.south east)+(0.5,0)$) -- ++(0,6) node (l1) {};
  \draw[loosely dashed] ($(itool-c.south west)+(-0.5,0)$) -- ++(0,6) node (l2) {};
  
  \node at (l1-|itool) {\normalsize Model Owner};
  \node at ($(l1)!0.5!(l2)$) {\normalsize Public};
  \node at (l2-|itool-c) {\normalsize Data Owner};
  
  \node[stepnum] at (keyspec.south east) {1};
  \node[stepnum] at (keygen-c.south east) {2};
  \node[stepnum] at (encrypt-c.south east) {3};
  \node[stepnum] at (inf.south east) {4};
  \node[stepnum] at (decrypt-c.south east) {5};
  
\end{tikzpicture}
    \caption{Functional schematic of \HEMAN{} including a model owner and a data owner.}
    \label{fig:architecture}
\end{figure*}
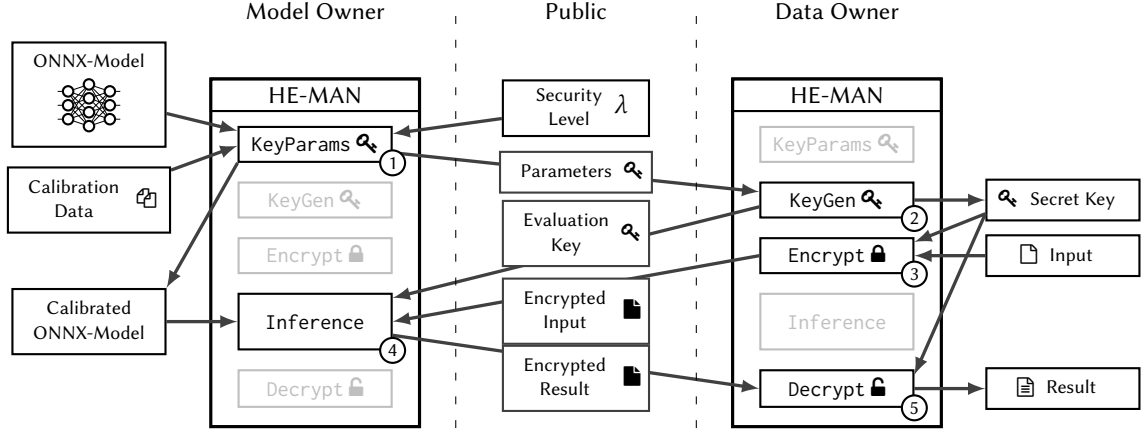
\section{Threat Model}\label{sec:threat-model}
We assume a semi-honest (also called honest-but-curious) threat model as in~\cite{juvekar2018gazelle,liu2017oblivious}, i.e. the parties are guaranteed to run the protocol and its computations faithfully, however model owner and data owner try to infer private information about the input data or model, respectively.
As the model owner only processes data encrypted using FHE schemes based on the assumed hardness of the ring learning with errors (RLWE) problem~\cite{lyubashevsky2010ideal}, privacy of the data owner is sufficiently ensured.
On the other hand, the model is never transferred to the data owner. Therefore, no sensitive information, i.e. model weights or the model architecture, can be leaked. In the case of \HEMANtenseal{} the encryption parameters contain information about the length of the coefficient modulus which is derived from the model and the calibration data. As a consequence, the data owner can derive the maximum number of performed multiplications and the maximum domain of intermediate values during inference from these values. However, neither the exact model architecture nor any weight values are leaked. The encryption parameters in \HEMANconcrete{} do not contain information that leaks any sensitive information.

If inference should be performed on a cloud server, the server is also assumed to be semi-honest. The cloud server receives the model in encrypted form, which adds a penalty in terms of computational execution time. Besides that, the cloud server only operates on encrypted data, thus, it cannot infer any private information.
We note that model inversion~\cite{fredrikson2015model} and model extraction attacks~\cite{tramer2016stealing,papernot2017practical} are out of scope of this work.

\section[HE-MAN]{\HEMAN{} -- Homomorphically Encrypted Machine Learning with ONNX Models}\label{sec:he-man}
Our \toolset{} aims to extend the landscape of FHE tools, making it easier to incorporate FHE into ML applications without deep FHE knowledge. Thus, a design goal of \HEMAN{} is to abstract away FHE details from the users, e.g. encryption parameter selection or transformations of computations into FHE-friendly operations.
We focus on a two-party Machine Learning as a Service setting, with a model owner (or server) who wants to provide inference as a service to data owners (or clients), while preserving privacy of the model and client-data. We extend the high-level illustration of Figure~\ref{fig:firstlady} with technical details in Figure~\ref{fig:architecture}.

\subsection{Overview}
Figure~\ref{fig:architecture} is divided into three segments. On the left the model owner side, on the right the data owner side and in between a public segment. Data on the model owner side (ONNX model, calibration data) and the data owner side (cleartext input and result, secret key) is never transferred to the opposite party. Data that is illustrated in the public segment is exchanged between the parties and contains no (or only very limited) sensitive information, such that neither the model nor the inputs are leaked to the opposite party.
Both parties use the same implementation of \HEMAN{}, however depending on the role (model owner or data owner) different commands are executed (highlighted rectangles).

\subsection{Initialization and Parameter Derivation}
First, the model together with a calibration dataset and a specified security level, e.g. 128 bits, are used to derive encryption parameters on the server-side~(\tikz[baseline=-0.6ex]{\node[draw,circle,inner sep=0.5mm,fill=white,thick] {\scriptsize 1};}).
To address a broad spectrum of models, \HEMAN{} accepts models in the Open Neural Network Exchange (ONNX) format, which is widely supported across all major ML frameworks.
The calibration dataset is used to derive a heuristic measure of the domains, e.g. lower and upper bounds, of intermediate results in the model by passing the calibration data through the model. This is used to derive proper encryption parameters and to perform efficient ciphertext operations during inference. 
As the calibration dataset is used to derive the minimum and maximum value per intermediate value in the model, i.e. edge in the ONNX graph, the calibration dataset needs to be a representative dataset for the corresponding neural network inference task. The training dataset or a large enough subset are suitable choices for the calibration dataset.
The security level is defined in bits and specified on the one hand by the client's requirements and on the other hand by technical restrictions of the model owner.
The outputs are a parameter file used for the subsequent key generation and a calibrated ONNX model which is extended by domain-information for each edge in the model.
The encryption parameters are sent to the client and used to generate the secret key (for encryption and decryption) and the evaluation key (for computations on encrypted data) on the client-side~(\tikz[baseline=-0.6ex]{\node[draw,circle,inner sep=0.5mm,fill=white,thick] {\scriptsize 2};}).

\subsection{Privacy-Preserving Inference}
Next, the client encrypts the input data using the secret key~(\tikz[baseline=-0.6ex]{\node[draw,circle,inner sep=0.5mm,fill=white,thick] {\scriptsize 3};}).
The encrypted input is sent to the server who performs inference using the calibrated model and the evaluation key~(\tikz[baseline=-0.6ex]{\node[draw,circle,inner sep=0.5mm,fill=white,thick] {\scriptsize 4};}). The evaluation key can be reused by the server for subsequent inference tasks to reduce communication overhead.
Finally, the encrypted result is returned to the client who decrypts it using the secret key~(\tikz[baseline=-0.6ex]{\node[draw,circle,inner sep=0.5mm,fill=white,thick] {\scriptsize 5};}) to obtain the cleartext result.
Steps~\tikz[baseline=-0.6ex]{\node[draw,circle,inner sep=0.5mm,fill=white,thick] {\scriptsize 1};} and~\tikz[baseline=-0.6ex]{\node[draw,circle,inner sep=0.5mm,fill=white,thick] {\scriptsize 2};} have to be executed only once before the first inference. Subsequent computations only comprise encryption~(\tikz[baseline=-0.6ex]{\node[draw,circle,inner sep=0.5mm,fill=white,thick] {\scriptsize 3};}), inference~(\tikz[baseline=-0.6ex]{\node[draw,circle,inner sep=0.5mm,fill=white,thick] {\scriptsize 4};}), and decryption~(\tikz[baseline=-0.6ex]{\node[draw,circle,inner sep=0.5mm,fill=white,thick] {\scriptsize 5};}) as the precomputed keys can be reused.

\subsection{Security Considerations}
The server is unable to gain any information about the client's data as only encrypted values are processed. Thus, the security of the data owner's private information is directly connected to the security of the underlying FHE scheme.
Furthermore, the model never leaves the model owner's region of trust. Therefore, the data owner is not able to infer the architecture or the specific parameters of the model. The encryption parameters are derived from the model. In the case of \HEMANtenseal{} they include information about the maximum number of multiplications and the maximum value domain in the model. In \HEMANconcrete{} they include expected value ranges for the input, as observed from the calibration dataset.

\section{Design Choices}\label{sec:designchoices}
In this section, we justify the choices we made during the design of \HEMAN{}.
\subsection{Command Line Interface}
Operations \tikz[baseline=-0.6ex]{\node[draw,circle,inner sep=0.5mm,fill=white,thick] {\scriptsize 1};} to \tikz[baseline=-0.6ex]{\node[draw,circle,inner sep=0.5mm,fill=white,thick] {\scriptsize 5};} illustrated in Figure~\ref{fig:architecture} and described in section~\ref{sec:he-man} can be executed via a user-friendly command line interface. For instance, the following command runs the \texttt{KeyParams} operation in \HEMANtenseal{}:\\
\texttt{tenseal-inference keyparams -m model.onnx -c calibration-data.zip -o keyparams.json}\\
\texttt{KeyParams} takes three parameters: the input ONNX model (\texttt{-m}), the calibration data file (\texttt{-c}) and the output encryption parameter file which will be generated. By default, parameters for a security level $\lambda=128$~bits are generated. We refer to the code repository for further information about the other commands (see Appendix~\ref{appendix:code}).

\subsection{Model Input via ONNX}
\begin{wrapfigure}{R}{0.5\textwidth}
    \centering
    \begin{lstlisting}[
        language=Python,
        caption=Example ONNX export in Python,
        label=lst:onnx,
        basicstyle=\ttfamily\tiny,
        numbers=left,
        numbersep=-10pt,
    ]
    import torch
    import torchvision
    
    model = torchvision.models.resnet50(pretrained=True)
    dummy_input = torch.empty(1, 3, 224, 224)
    torch.onnx.export(model, dummy_input, "resnet50.onnx")
    \end{lstlisting}
\end{wrapfigure}
\noindent\HEMAN{} features ONNX as model input format, as it is supported either directly or by dedicated conversion tools for all major ML libraries. This constitutes the ability to use a broad range of pretrained networks. Still, ONNX models are easy to integrate due to \mbox{\emph{Protobuf}}\footnote{\url{https://developers.google.com/protocol-buffers/}} being the basis of ONNX which is widely supported across programming languages.
ONNX uses a directed graph representation for neural networks. Nodes correspond to applications of operators, e.g. a network layer or an activation function, edges represent how data and (intermediate) results flow through the network. Listing~\ref{lst:onnx} contains example code for loading a pretrained ResNet50 model and exporting in the ONNX format. Unlike earlier efforts, \HEMAN{} provides out-of-the-box support for a wide variety of ML models in the ONNX format.

\subsection{Encryption Parameter Selection}
\begin{wraptable}{r}{0.35\textwidth}
\vspace{-5mm}
\centering
\caption{\HEMAN{} Encryption Parameters for \concrete{} (top) and TenSEAL (bottom)}\label{tab:encryption_params}
    \begin{tabular}{ccccc}
      \multicolumn{5}{c}{\concrete{}}\\
      \toprule
       & \multicolumn{2}{c}{${\text{RLWE}}$} & \multicolumn{2}{c}{${\text{LWE}}$} \\ \cmidrule(lr){2-3}\cmidrule(lr){4-5}
      $\lambda$ & $N$ & $\sigma$ & $k$ & $\sigma$ \\ \midrule
      80 
       & 2048 & $2^{-60}$ & 542 & $2^{-23}$ \\ \midrule
      128 & 4096 & $2^{-62}$ & 938 & $2^{-23}$ \\
      \bottomrule\\
    \end{tabular}
    \begin{tabular}{rcc}
    \multicolumn{3}{c}{TenSEAL}\\
    \toprule
      & & \scriptsize$\lambda=128$ \\ \cmidrule(lr){3-3}
      $N$ & $\log_2 N$ & $\log_2 q$\\
      \midrule
      4096 & 12 & 109  \\
      8192 & 13 & 218  \\
      16384 & 14 & 438 \\
      32768 & 15 & 881 \\
      \bottomrule
    \end{tabular}
    \vspace{-4mm}
\end{wraptable}
One of the main tasks of \HEMAN{} is to automatically choose encryption parameters such that users are not required to have expert knowledge of FHE.

Given a neural network and a security level, the encryption parameters are selected as small as possible to maximize performance while still guaranteeing accurate computation and satisfying the specified security level, e.g. 128 bits. Table~\ref{tab:encryption_params} lists the encryption parameter sets implemented in \HEMAN{}.

In the case of \HEMANconcrete{}, parameters based on~\cite{chillotti2020programmable} are used, guaranteeing enough bits of precision after bootstrapping and ensuring the security level. 

The parameter selection for \HEMANtenseal{} is affected mostly by the maximum multiplicative depth of the computation, i.e. the depth of the neural network.
For optimal performance, the polynomial modulus degree~$N$ is chosen as small as possible, such that the coefficient modulus~$q$ is large enough to be able to perform the required amount of multiplications with adequate precision. Pairs of $(N,q)$ yielding certain security limits are implemented according to~\cite{HomomorphicEncryptionSecurityStandard,laine2017simple}.
For details about the encryption parameter selection, we refer to the crypto module in the source-code. 

\subsection{Homomorphic Inference}
After encrypting the input with a secret key based on the derived parameters, \HEMAN{} allows to compute the model forward pass.
By implementing homomorphic versions of several of the ONNX specification operator set, our tools provide a homomorphic runtime for a subset of ONNX neural networks.
We currently support elementwise additions and multiplications, matrix multiplications, convolutions, average pooling, ReLU activations, and padding.
Depending on the library backend, we further support a number of operators implementing different activation functions.

So far, we do not apply optimizations directly to the ONNX computational graph. 
The homomorphic runtime does however keep an internal state over the course of a forward pass to allow efficient execution.
Operators are executed as batched or contracted operations where possible in order to improve efficiency.

\subsection[HE-MAN-Concrete]{\HEMANconcrete{}}
\noindent
\textbf{Heuristic ciphertext intervals:}
To guarantee correct decryption in \concrete{}, it is required to track the interval in which each encrypted value falls. 
The interval of an operator's result can be bounded analytically by considering the interval of the operands. 
In practice however, this leads to an overestimation of the interval and degrades the accuracy via greater quantization errors.
We instead use the calibration data during the \texttt{KeyParams} step~(\tikz[baseline=-0.6ex]{\node[draw,circle,inner sep=0.5mm,fill=white,thick] {\scriptsize 1};} in Fig.~\ref{fig:architecture}) to trace the intervals of values at all stages of the network.
A cleartext forward pass through the network is performed using the calibration data to determine heuristic interval bounds which are stored in the calibrated model and are more favorable than analytic worst-case bounds.
\\ \\
\textbf{Bootstrap folding:}
\HEMANconcrete{} performs bootstrapping frequently to evaluate non-linear univariate functions or to refresh ciphertext noise and padding.
We do not run the bootstrapping algorithm immediately, but instead push it on a stack of operations unique to each tensor.
Evaluation of the stack is deferred until a point in the computational graph where any non-univariate operator depends on the result.
This allows folding of multiple sequential bootstraps into a single operation, significantly improving runtimes.

\subsection[HE-MAN-TenSEAL]{\HEMANtenseal{}}
\textbf{Arbitrary linear operations:} 
Linear operations, e.g. convolutions or general matrix multiplications, are realized in \HEMANtenseal{} using vector-matrix multiplications between the batched ciphertext vector and a weight matrix with an optional bias addition. This results in an increased number of multiplications and therefore longer execution time compared to more efficient techniques leveraging ciphertext rotations. However, this allows for arbitrary linear operations in the model regarding number and position of the operations, in contrast to the TenSEAL API, that only allows a single convolutional layer. Moreover, a convolution has to be the first operation in the network and the inputs to this layer have to be encrypted (by the data owner) using a dedicated encryption function, hence, partially leaking the architecture to the data owner.
\\ \\
\begin{wrapfigure}{r}{0.4\textwidth}
  \begin{center}
    \input{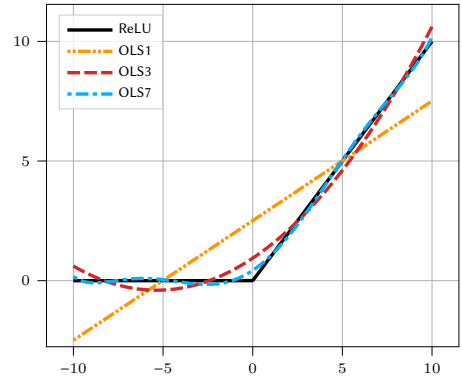}
  \end{center}
  \caption{Examples of polynomial approximations of the ReLU function for different approximation methods in the interval [-10,10]. Best viewed in color.}
  \label{fig:relus}
\end{wrapfigure}
\noindent\textbf{ReLU approximation: }
As TenSEAL only allows for polynomial computations, nonlinear activations need to be approximated by polynomials. Previous works use square activations~\cite{gilad2016cryptonets,aharoni2020helayers} or polynomials~\cite{dathathri2020eva} where the coefficients are learned during training. However, when computing a forward pass using pretrained models with ReLU activations, this is not applicable. 

As CKKS ciphertexts contain vectors of encrypted values, one ciphertext holds all values of an edge in the ONNX graph, i.e. all values per layer in the network. To compute optimal ReLU approximations we analyze the input values of ReLU operations per layer in the network.
Similar to \HEMANconcrete{}, we use the calibration data for a forward pass through the network within the \texttt{KeyParams} step to determine the domain for a polynomial to approximate the ReLU activation.
We implemented two methods to calibrate the approximation interval. First, the interval is determined by the minimum and maximum value within the layer, i.e. $[\text{min}, \text{max}]$. In our second approach, we compute the mean $\mu$ and the standard deviation $\sigma$ of the ReLU input values and set the approximation domain to $[\mu-3\sigma, \mu+3\sigma]$.
This domain is narrower for normally distributed input values, which is always the case in our experiments. Hence, some values lie outside of the approximation domain, but a large portion of values (around the mean) exhibit a smaller approximation error. 
We evaluate both domain calibration methods on three ordinary least squares polynomial approximations of degree one (OLS1), three (OLS3), and seven (OLS7). These are polynomials with maximum degree and a multiplicative depth of one, two, and three, respectively. Recall, higher multiplicative depth requires a higher polynomial modulus degree $N$, resulting in increased execution time. We measure the accuracy on the full MNIST~\cite{lecun2010mnist} test set of 10\,000 samples using a LeNet-5 network that has been trained with exact ReLU activation functions.
Figure~\ref{fig:relus} illustrates examples of approximations along with the exact ReLU function.

Table~\ref{tab:relu} summarizes the results. For every method we evaluate the classification accuracy for both interval calibration methods and state the multiplicative depth $d_m$ for a single ReLU approximation as well as the network's required polynomial degree~$N$.
OLS1 yields unusable accuracy, which implies that a polynomial with minimum multiplicative depth of two is necessary.
OLS3 gives reasonable accuracies, particularly using the \meanstd{} calibration method.
OLS7 with the min-max intervals gives the best accuracy. The substantial accuracy drop for the \meanstd{} method comes from the fact that polynomials with high degree result in extremely large (absolute) values outside of the approximation interval. Therefore, values outside of the approximation interval dominate the subsequent computation, resulting in unpredictable behavior.
For the following evaluation, OLS3 \meanstd{} (highlighted in Table~\ref{tab:relu}) is used to approximate ReLU operators in \HEMANtenseal{} as this yields the best tradeoff between accuracy and multiplicative depth.
\begin{table}
\centering
\parbox{.4\linewidth}{
\caption{Comparison of ReLU approximations in TenSEAL using polynomials}
\resizebox{\linewidth}{!}{%
\centering
    \begin{tabular}{llcccc}
        \toprule
        \multicolumn{2}{l}{\textbf{Network}} & \multicolumn{2}{c}{Calibration} & & \\ \cmidrule(lr){3-4}
        & Method & min - max & $\mu \pm 3\sigma$ & $d_m$ & $\log_2 N$ \\
        \midrule
        \multicolumn{2}{l}{\textbf{LeNet-5}} &&&&\\
        & OLS1 & .097 & .097 & 1 & 14\\
        & OLS3 & .872 & \cellcolor{gray}\textbf{.954} & 2 & 14\\
        & OLS7 & .985 & .151 & 3 & 15\\
        \bottomrule
    \end{tabular}\label{tab:relu}}
}
\qquad
\parbox{.5\linewidth}{
\caption{Neural Networks used in the evaluation}
\resizebox{\linewidth}{!}{%
\centering
    \begin{tabular}{llccccc}
        \toprule
        \multicolumn{2}{l}{\textbf{Dataset}} & \multicolumn{3}{c}{No. of layers} & & \\ \cmidrule(lr){3-5}
        &Network & Conv & FC & ReLU & $p$ & accuracy \\ \midrule
        \multicolumn{2}{l}{\textbf{MNIST}} & & & & & \\
        &CryptoNets & 1 & 2 & 2 & 52722 & .975 \\
        &LeNet-5 & 3 & 2 & 4 & 61706 & .991 \\
        \midrule
        \multicolumn{2}{l}{\textbf{LFW}} & & & & & \\
        & MobileFaceNets & \multirow{2}{*}{2} & \multirow{2}{*}{0} & \multirow{2}{*}{0} & \multirow{2}{*}{56960} & \multirow{2}{*}{.990}\\ 
        & (classifier) & & & & & \\ \bottomrule
    \end{tabular}\label{tab:networks}}
}
\end{table}

\section{Evaluation}\label{sec:evaluation}
To assess the feasibility of our approach, we evaluate the accuracy and inference time of \HEMAN{} as well as the automatic choice of encryption parameters. Our task domains comprise handwritten digit classification and face recognition. Table~\ref{tab:networks} gives an overview of the networks used during evaluation, including the number of convolutional layers (Conv), fully connected layers (FC) and ReLU activations (ReLU), as well as the number of parameters $p$ in the pretrained model and the accuracy using cleartext data. Detailed information about the architectures can be found in Appendix~\ref{appendix:networks}. We evaluate cleartext accuracy using the full 10\,000 test samples for MNIST and 6\,000 samples for the face recognition task.
\subsection{Setup}
We use the MNIST dataset of handwritten digits~\cite{lecun2010mnist} to train one architecture of CryptoNets~\cite{gilad2016cryptonets} and one \mbox{LeNet-5}~\cite{lecun1998gradient}. Inputs are padded with zero values such that they match the input shape of 32\,$\times$\,32 from LeNet-5. All activation functions are set to ReLU.

For the face recognition task we use the MobileFaceNets architecture~\cite{chen2018mobilefacenets} together with the CASIA Webface dataset for training and the Labeled Faces in the Wild (LFW) dataset for evaluation. Example images from both datasets are illustrated in Appendix~\ref{appendix:faces}. The final activation function was changed from PReLU to ReLU, in order to be amenable to homomorphic execution. The number of channels in the second to last convolutional layer was reduced in order to fit the resulting tensor into a CKKS ciphertext with polynomial degree $N$ of~$2^{15}$. Training data consists of approximately 490k cropped and aligned RGB images of faces of size 112\,$\times$\,112\,$\times$\,3. The model outputs a 128-dimensional feature vector for an input face. We train the network using arcface loss~\cite{deng2019arcface}.
Test accuracy is evaluated using a different dataset, namely the Labeled Faces in the Wild (LFW). For a successful test case the model needs to correctly decide whether two input images depict the same person.
If the cosine similarity of the two feature vectors exceeds a certain threshold they are considered as originating from the same person.
The authors of~\cite{chen2018mobilefacenets} report a test accuracy of 99.55~\%. With the described modifications to the architecture we achieve a slightly diminished test accuracy of 99.00~\% in the clear.

As the entire MobileFaceNets architecture is too complex to be usable in the homomorphic domain, we use an approach inspired by~\cite{schlogl2020ennclave}. Here, the authors split the model into an ``extractor'' and a ``classifier''\footnote{Note that this does not directly correspond to the extractor and classifier parts as typically used in transformer-based ML models.} part, where the latter is executed in a trusted enclave and the ``extractor'' forward pass can be computed locally at the client-side. We transfer this approach into the FHE domain and split the MobileFaceNets architecture also into an ``extractor'' and ``classifier'' part. The ``extractor'' on its own is not sufficient to perform the face recognition task and is computed locally by the data owner. The outputs of this part are then encrypted and sent to the model owner who executes the ``extractor'' part which does not need to be disclosed to the data owner.

All experiments were run on an AMD Ryzen 7 2700X 8-core CPU with 64 GB of RAM (running Kubuntu 20.04). Experiments can be rerun using the evaluation code along with the source-code of \HEMAN{}. 

\subsection{Results}
\begin{table}
\centering
\parbox{.4\linewidth}{
\caption{Encryption Parameters selected by \HEMANtenseal{} yielding 128 bits of security}
\centering
    \begin{tabular}{llccc}
        \toprule
        \multicolumn{2}{l}{\textbf{Dataset}} & \multirow{2}{*}{$\log_2 N$} & \multirow{2}{*}{$\log_2 q$} & \multirow{2}{*}{$d_m$} \\
        & Network & & & \\ \midrule
        \multicolumn{2}{l}{\textbf{MNIST}} & & & \\
        & CryptoNets & 13 & 218 & 7 \\
        & LeNet-5 & 14 & 437 & 15 \\
        \midrule
        \multicolumn{2}{l}{\textbf{LFW}} & & & \\
        & MobileFaceNets & \multirow{2}{*}{15} & \multirow{2}{*}{228} & \multirow{2}{*}{2} \\
        & (classifier) & & & \\
        \bottomrule
    \end{tabular}\label{tab:parameters}}
\qquad
\parbox{.55\linewidth}{
\vspace{4.5mm}
\caption{Evaluation Results}
\resizebox{\linewidth}{!}{%
\centering
    \begin{tabular}{llcrcr}
        \toprule
        \multicolumn{2}{l}{\textbf{Dataset}} & \multicolumn{2}{c}{\HEMANconcrete{}} & \multicolumn{2}{c}{\HEMANtenseal{}}\\ \cmidrule(lr){3-4} \cmidrule(lr){5-6}
        & Network & accuracy & latency & accuracy & latency\\ \midrule
        \multicolumn{2}{l}{\textbf{MNIST}} & & & & \\
        & CryptoNets & .968 & 112 s & .924 & 8 s \\
        & LeNet-5 & .984 & 1672 s & .789 & 236 s \\
        \midrule
        \multicolumn{2}{l}{\textbf{LFW}} & & & & \\
        & MobileFaceNets & \multirow{2}{*}{.970} & \multirow{2}{*}{68 s} & \multirow{2}{*}{.972} & \multirow{2}{*}{196 s} \\ 
        & (classifier) & & & & \\ \bottomrule
    \end{tabular}\label{tab:results}}
}
\end{table}

Our final results are shown in Tables~\ref{tab:parameters} and~\ref{tab:results}.
Table~\ref{tab:parameters} lists the encryption parameters that are derived in \HEMANtenseal{} and used for evaluation.
Our experiments show similar results to~\cite{dathathri2020eva} for comparable networks (LeNet-5).
\HEMANconcrete{} always uses the 128 bit security parameter set listed in Table~\ref{tab:encryption_params}.

Table~\ref{tab:results} shows the accuracies and latencies, i.e. the mean inference time for a single sample, for both tasks and all networks. Due to long execution times we evaluate all homomorphic inference experiments using 1\,000 samples per experiment.
\HEMANtenseal{} exhibits accuracies of 92.4~\% and 78.9~\% for MNIST with the CryptoNets and LeNet-5 architecture, respectively. While accuracy only drops slightly for CryptoNets, there is a significant decrease with LeNet-5, which is caused by the selection of encryption parameters.
Accuracy could be improved using encryption parameters with \mbox{$N = 2^{15}$} at the cost of increased execution time.

Alternatively, accuracy could be improved by using the same polynomial activation function during training and inference, e.g. a square activation. However, as discussed above, models with polynomial activations are hardly ever used.

\HEMANconcrete{} shows accuracies of 96.8~\% and 98.4~\% for CryptoNets and LeNet-5, respectively. This only slightly decreases cleartext accuracy (cf.~Table~\ref{tab:networks}). This comes at the cost of increased latency by a factor of 14.0 and 7.1 compared to \HEMANtenseal{} for CryptoNets and LeNet-5, respectively.

Latency of MobileFaceNets is much lower for \HEMANconcrete{} with 69 s than for \HEMANtenseal{} which exhibits 208 s. Without any nonlinear activation function in the network, bootstrapping is not necessary to compute the forward pass, thus reducing inference time drastically.
The accuracies of both tools, 97.0~\% for \HEMANconcrete{} and 97.2~\% for \HEMANtenseal{}, is on par with the corresponding accuracy using cleartext input data.

Regarding execution time we heavily depend on the implementation of open-source homomorphic encryption libraries. Any improvement of a backend library would improve execution times of \HEMAN{}. It is easy to extend \HEMAN{} with additional HE libraries, as exemplified by our implementation using \concrete{} and TenSEAL.

\section{Related Work}\label{sec:related-work}
\subsection{Privacy-Preserving Deep Learning}
CryptoNets~\cite{gilad2016cryptonets}, the first demonstration of neural network inference using FHE, initiated subsequent developments, e.g. supporting more complex networks~\cite{brutzkus2019low} or efficient inference using discretized neural networks with sign activation function~\cite{bourse2018fast}.
Secure inference frameworks emerged using Secure Multi-Party Computation (SMPC)~\cite{riazi2019xonn, ganesan2022efficient}, or following a hybrid approach (FHE and SMPC)~\cite{juvekar2018gazelle, rathee2020cryptflow2, knott2021crypten, lehmkuhl2021muse, huang2022cheetah, mishra2020delphi}.
CrypTFlow2~\cite{rathee2020cryptflow2} and CrypTen~\cite{knott2021crypten} are frameworks that provide easy to use APIs that integrate with TensorFlow and PyTorch, respectively. While these frameworks are solid solutions for their respective models, \HEMAN{} only requires an ONNX representation of the model and is therefore independent of any ML framework.
Gazelle~\cite{juvekar2018gazelle}, Delphi~\cite{mishra2020delphi}, Cheetah~\cite{huang2022cheetah} and Muse~\cite{lehmkuhl2021muse} switch between SMPC and FHE primitives based on the alternating structure of linear and non-linear layers in neural networks. The protocol of Muse~\cite{lehmkuhl2021muse} is secure against malicious clients.
These SMPC approaches are usually more efficient than their pure FHE counterparts. For instance Cheetah evaluates a MiniONN network, which is similar in size to the LeNet-5 network, in 3.55 seconds and a ResNet50 network in 80.3 seconds. 
However, SMPC protocols exhibit a substantial communication cost, e.g. for Cheetah 30 MB and 2.3 GB of communication for the above examples, respectively~\cite{huang2022cheetah}.
Communication cost of \HEMAN{} comprises of the exchange of encrypted inputs and encrypted results. For example, private inference for one MNIST sample consumes 8 MB and 2 MB of communication for \HEMANconcrete{} and \HEMANtenseal{}, respectively. This does not include the initial transfer of the evaluation key as this has to be done only once between a model owner and a data owner for any number of subsequent evaluations. Additionally, larger networks only slightly increase communication cost contrary to SMPC protocols.
SMPC also requires the joint computation to be known by all parties, hence leaking potential intellectual property of the model owner, e.g. the model architecture to the data owner.

\subsection{General Purpose FHE Compilers}
As writing FHE programs requires substantial domain knowledge, general purpose compilers, also called FHE compilers, that translate standard code into FHE programs have been proposed. Compilers targeting different programming languages and FHE schemes~\cite{carpov2015armadillo,crockett2018alchemy} were implemented. Recently, HECO was proposed, an end-to-end FHE compiler introducing a new compiler design, extending the scope of optimizations of previous works~\cite{viand2022heco}.

\subsection{FHE Compilers for ML}
The nGraph-HE compiler~\cite{boemer2019ngraph} and its second version~\cite{boemer2019ngraph2} were one of the first compilers designed for ML applications and enable TensorFlow~\cite{tensorflow2015-whitepaper} model inference over encrypted data. However, encryption parameters must be manually set by the user, thus, missing an important piece of abstraction for non-experts.
Subsequent compilers solved this issue through automatic encryption parameter selection. SEALion~\cite{van2019sealion} exhibits an easy to use high-level API for encrypted inference of neural networks using the BFV~\cite{fan2012somewhat} scheme implemented in SEAL. Encryption parameters are derived using a heuristic search algorithm. Non-polynomial functions, e.g. ReLU or sigmoid, are not supported. SEALion is currently not publicly available.
%
The EVA compiler~\cite{dathathri2020eva} is built on top of the author's prior work CHET~\cite{dathathri2019chet} utilizing the CKKS scheme in SEAL.
While being designed for neural network inference, EVA is capable of encrypted computations of any application.
The encryption parameters are derived from the computation circuit to maximize performance while ensuring security and correctness.
While \HEMAN{} only requires the neural network in the ONNX format, EVA needs the function to be defined using a newly introduced input language. Moreover, activation functions must be replaced by polynomials. EVA is publicly available, CHET is currently not.
Importing neural networks via a standard file format is supported by HeLayers~\cite{aharoni2020helayers} and PlaidML-HE~\cite{chen2019plaidml}. In contrast to \HEMAN{}, both are evaluated only using the CKKS scheme implemented in SEAL. Hence, nonlinear operations are replaced by FHE-friendly functions, e.g. square activations.

\section{Conclusion}\label{sec:conclusion}
Machine learning services handling sensitive data require particular attention to ensure privacy of data owners.
FHE is a suitable technique as service providers only have access to encrypted data.
However, ML applications rarely make use of FHE as implementing efficient and secure applications requires expertise in the underlying cryptographic methods.

In this work we introduce \HEMAN{}, a \toolset{} for neural network inference on homomorphically encrypted data. 
Taking a trained model in the ONNX format and an encrypted input, our tools compute the forward pass homomorphically without requiring network-specific optimization, or hand-selection of cryptographic parameters.
By implementing tools using \concrete{} and TenSEAL, we support two widely used FHE libraries that offer suitable properties depending on the use case.
\HEMAN{} is specifically geared towards the setting where a service provider has an expensively trained model and wants to provide inference as a service, keeping both the model weights and architecture, as well as the customer's input data private.
Service providers can offer their inference service on homomorphically encrypted data without any extra effort considering FHE. To sum up, this enables various sensitive ML applications without compromising privacy of either party.

\HEMAN{} requires a pretrained model as an input. Training such a model on encrypted data is possible for simple ML models~\cite{graepel2012ml,bergamaschi2019homomorphic} but, as of now, not applicable to complex structures such as neural networks in practical execution times~\cite{nandakumar2019towards} or without major restrictions~\cite{aono2017privacy}.

One avenue for future development entails the support for a larger set of operators from the ONNX specification, with the eventual goal of support for the full format. 
Another promising direction would see the unification of our tools under a common frontend, as well as support for further homomorphic backend libraries. 
Since each scheme has its strengths and drawbacks, this would allow users to choose the best fit for their particular use case.

Finally, we want to highlight that technical developments, such as parallelization support, e.g. with GPUs, of the backend libraries implemented in \HEMAN{} would instantly boost the efficiency of our \toolset{}. The same holds true for other improvements and optimizations and we invite other researchers and practitioners to extend our \toolset{} with additional FHE libraries. This could on the one hand further improve the latency and on the other hand diversify the FHE schemes supported by \HEMAN.

\begin{acks}
All authors of this work are supported under the project ``Secure Machine Learning Applications with Homomorphically Encrypted Data'' (project no. 886524) by the Federal Ministry for Climate Action, Environment, Energy, Mobility, Innovation and Technology (BMK) of Austria.
\end{acks}

\bibliographystyle{ACM-Reference-Format}
\bibliography{main}

\appendix

\section{Neural Network Architecture}\label{appendix:networks}
Table~\ref{tab:architectures} shows the detailed architecture of the networks involved during evaluation.

\section{Source-Code Access}\label{appendix:code}
Source-code of \HEMANconcrete{} is publicly available at \texttt{https://github.com/smile-ffg/he-man-concrete} and of \HEMANtenseal{} at \texttt{https://github.com/smile-ffg/he-man-tenseal}.
Further information on how to execute \HEMAN{} or how to perform the evaluations can be found in readme files within the repository.

\section{Face Recognition Dataset Examples}\label{appendix:faces}
Figure~\ref{fig:faces} shows examples from the face recognition datasets. Four examples of the CASIA Webface dataset which is used for training are illustrated in Figure~\ref{fig:faces}(a). Evaluation dataset examples from the Labeled Faces in the Wild (LFW) dataset are illustrated in Figure~\ref{fig:faces}(b).
Note that the two rightmost images in Figure~\ref{fig:faces}(b) depict the same person and are inputs for the evaluation, i.e. the evaluation is successful if the face detector recognizes the two images as originating from the same person.

\begin{table}[hb!]
\parbox{.50\linewidth}{
\caption{Architecture of the networks used in the evaluation.}
\resizebox{\linewidth}{!}{%
\centering
    \begin{tabular}{l l c c c c c}
        \toprule
        \multicolumn{2}{l}{\textbf{Network}} & \multicolumn{2}{c}{channels} & \multicolumn{2}{c}{kernel} & \\
        \cmidrule(lr){3-4}\cmidrule(lr){5-6}
        No. & Layer & in & out & size & stride & activation \\ \midrule
        \multicolumn{7}{l}{\textbf{CryptoNets}} \\
        1 & Conv & 1 & 4 & (5,5) & 3 & ReLU\\
        2 & FC & 400 & 128 & & & ReLU\\
        3 & FC & 128 & 10 & & &\\
        \midrule
        \multicolumn{5}{l}{\textbf{LeNet-5}} \\
        1 & Conv & 1 & 6 & (5,5) & 1 & ReLU\\
        2 & AveragePool & 6 & 6 & (2,2) &  & \\
        3 & Conv & 6 & 16 & (5,5) & 1 & ReLU\\
        4 & AveragePool & 16 & 16 & (2,2) & & \\
        5 & Conv & 16 & 120 & (5,5) & 1 & ReLU\\
        6 & FC & 120 & 84 & & & ReLU\\
        7 & FC & 84 & 10 & & &\\
        \midrule
        \multicolumn{7}{l}{\textbf{MobileFaceNets (classifier)}} \\
        1 & DepthwiseConv & 320 & 320 & (7,7) & 1 & \\
        2 & Conv & 320 & 128 & (1,1) & 1 & \\
        \bottomrule
    \end{tabular}\label{tab:architectures}}
}
\hfill
\parbox{.45\linewidth}{
    \centering
    \begin{tikzpicture}
          \node[inner sep = 0] (1) at (0,0) {\includegraphics[scale=0.4]{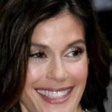}};
          \node[inner sep = 0, right=2mm of 1] (2) {\includegraphics[scale=0.4]{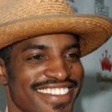}};
          \node[inner sep = 0, right=2mm of 2] (3) {\includegraphics[scale=0.4]{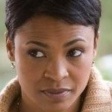}};
          \node[inner sep = 0, right=2mm of 3] (4) {\includegraphics[scale=0.4]{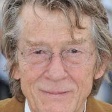}};
          \node at ($(2)!0.5!(3)+(0,-1.2)$) {\footnotesize (a) CASIA Webface examples};
          
          \node[inner sep = 0, below=1cm of 1] (5) {\includegraphics[scale=0.4]{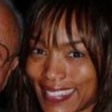}};
          \node[inner sep = 0, right=2mm of 5] (6) {\includegraphics[scale=0.4]{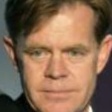}};
          \node[inner sep = 0, right=2mm of 6] (7) {\includegraphics[scale=0.4]{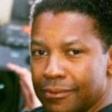}};
          \node[inner sep = 0, right=2mm of 7] (8) {\includegraphics[scale=0.4]{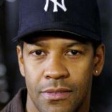}};
          \node at ($(6)!0.5!(7)+(0,-1.2)$) {\footnotesize (b) LFW examples};
    \end{tikzpicture}
    \captionof{figure}{Examples of images used for training and evaluation of the face recognition task.}
    \label{fig:faces}
}
\end{table}

\end{document}